**Brain MRI detection by Sematic Segmentation models- Transfer Learning approach**

Jayanthi Vajiram [1], Aishwarya Senthil[2]

[1]Research Scholar (Jayanthi.2020@vitstudent.ac.in), Vellore Institute of Technology,

[2] MBBS Chettinad Medical Research Institute, (aishwarya.senthil00@gmail.com)

**Abstract**

The MRI is a Neuroimaging method in which various body organs and tissues are explained in previous studies. Segmentation techniques using a dataset of MR images to generate an accurately predicted mask is the paper outline. Automatic segmentation of an atrophic structure based on its dimensions and shape is compared with the mask and the predicted mask. For this approach, the transfer learning-based pre-trained model layers are used to do the new trainable layer. The pre-trained models are used to classify and segment glioma induced Brain Tumors based on its variations in the brain region. The segmentation of the affected region and its performance metrics helps us to detect and find the abnormality in the Brain. The convolutional neural network (CNN) is most widely used for the automatic segmentation of predicting the risk of Brain Tumor occurrence. The non-isotropic resolution, Rician noise, and bias field effects in MRI can't handle by machine learning accurately. Hence, these challenges, are overcome by merging and automating the relevant extracted features with the classification procedure of deep learning approaches. The proposed models like VGG16, ResNet50 and ResU-net is used to predict the MRI image through classification and segmentation problems based on the original mask and predicted image mask. The result shows the ResNet50 is a promising model based on the accuracy comparison and high F1-score of 0.82. The results are arrived at by the semantic segmentation method, which gives the spatial uncertainty maps and quantifies a prediction of the Brain Tumor by an image level, which helps the health professionals to find the specific cases and consider them for better clinical diagnosis and treatment.

**1 Introduction**

The magnetic resonance images used to detect the abnormalities in the various body organs and tissues are explained in previous studies (Rutegard,2017) and mainly used to analyze the anatomical structure of brain, in which the image segmentation and extracting features from the original images are processed for clinical diagnosis purpose. The classical approaches like region, boundary, hybrid and atlas-based approaches (Lin,2016). The neural network can easily handle bulky datasets. Automatic detection of Brain is a challenging step, due to its orientation, size and intensities. The MRI brain images are classified, due to the complexity of the brain tissue, different models have been developed to analyze different modalities and multimodal imaging techniques, which provides information about various tissue features of the brain. The various methods have been developed for Brain Tumor detection, including region growing methods (Deng,2010), edge-based methods (Aslam,2015), thresholding methods (Singh and Magudeeswaran,2017), morphological methods, atlas-based methods and fuzzy clustering technique. The challenges of previous methods in the detection of Brain Tumor, has noise and artifacts interruptions. In this methodology, an effective approach is used to minimize these challenges by implementing a transfer learning for the detection of the Brain Tumor from the MRI dataset [1]. Glioma is the main symptom for Brain Tumor and Seizure. The 80% of seizure induced by Brain Tumor and analyzed by multimodal MRI method [2].

ImageNet database has images and their labels. Pretrained models like VGG-16, ResNet50, and ResU-net are already trained by millions of images consisting of thousands of image categories. The model act as a feature extractor for new images of different categories from the source dataset, but the pre-trained model should still be able to extract relevant features from these images based on transfer learning. This paper explains the transfer learning by pre-trained models of VGG-16, ResNet50, and ResU-net as an effective feature extractor to classify and segment the Glioma-affected images even with fewer training images [3].

## 1.1 Related work

The accurate diagnosis and estimations are calculated by segmentation of the brain tissues. The MRI brain images of T1w, T2w, and PD-weighted images have different contrast characteristics of brain tissues (Akkus, 2017). The skull segmentation was proposed by Galdames in 2012. A new multi-atlas brain segmentation (mabs) method of comparison of atlas similarity to the target image (Del Re,2016). The Beast (brain extraction based on nonlocal segmentation technique) method was used to observe the brain mask (Manjon,2014). The level set method of gaussian distribution of brain extraction with region-based approach, which gives the better result proved Wang et al. (2010) [4].The deep learning models have made significant results in image classification (Rawat and Wang, 2017), segmentation (Garcia, 2017), image captioning (Anderson, 2018), object detection (Zhao, 2019), and object tracking (Tomasi,2018). The multi-modal analysis with image fusion of medical images gives better accuracy through the segmentation process (Cheng,2019). Christ et al. in [5] proposed two cascaded FCNs were employed to segment the ROI of the lesion by dense 3D CNN. Hamidian et al. converted 3D CNN with a fixed field of 3D FCN (Fully convolutional network) and fully trainable models with nonlinear mappings between inputs and outputs. This FCN is inspired by the VGG network structure [6]. Recently, transfer learning methods have shown promising performance in solving image classification, semantic segmentation, and detection [7]. In this paper, semantic segmentation models and their performances are used to study the Brain Tumor.ow-grade gliomas (LGGs) are a type of brain tumor that often arise at a young, the long-term treatment-related risks are more and cured by surgery, radiation and chemotherapy, and proper medications. Clinical symptoms vary from invasion to the presence of parenchyma or obstructive hydrocephalus. Seizure is the presenting symptom in up to 80% of LGG patients. It leads to behavioral or cognitive changes, focal neurologic deficits, headache, or papilledema in asymptomatic Brain tumor patients [8]. Survival possibilities by the year of diagnosis for Astrocytoma and all low-grade progress to high-grade glioma and death. Data from the Surveillance, Epidemiology, and End Results (SEER) are explained clearly in previous clinical studies [9]. Gliomas are brain tumors from glial cells. Gliomas can be low grade (slow growing) or high grade (fast-growing). Increased levels of *Myo-inositol* increase the glial cells, which leads to II gliomas and depends on the Choline level [10].

Parisot et al. used to classify the tumor first [11]. Huang et al. used the samples to build up a softmax model to optimize the different classes [12]. Meier et al. applied decision forests to discriminate pathological regions from brain MRI volumes [13]. Zhao et al. proposed a semi-segmentation [14], for label purposes. Meier et al. estimated the CRF of voxel-wise classification from a decision forest classifier [15]. Detection and classification of HGG and LGG brain tumors using machine learning based on the original and predicted mask of the tumor are tried in the earlier work [16].

Classifying tumors by using deep learning models, which use 3D volumetric MRIs or even 2D slices separately by two spatiotemporal methods, ResNet (2+1) D and ResNet mixed Convolution, to classify the types of brain tumors. Pre-trained ResNet Mixed Convolution gives a macro F1-score of 0.9345 and a test accuracy of 96.98% [17]. Since 2015–2016, medical applications use the deep learning models: cell segmentation by UNet [18], prostate segmentation by VNet [19], and 3D UNet [20]. The tumor segmentation methods were published in Miccai Brats Challenge [21]. The dataset for learning the U-net network for Brain Tumor Segmentation Challenge (BraTS). F-Score (0.669652 -HGG) was used to measure the accuracy of the learned network [22]. Gliomas are characterized by subtypes of HGG and LGG. The grade I tumor is removed by surgery, while grade IV tumors are difficult to treat [23-25]. The F-Score can be used to calculate more precisely [26].

Recently, there have been many developments in the segmentation of brain images, especially in brain tumor care using dual encoder structure and channel spatial attention block [27]. The model Resnet50 was introduced by Kaiming He and Jian Sun team in 2015 through 'Deep Residual Learning for Image Recognition. The skip connection concept is used to strengthen this 50-layer residual network model. It has the Keras pre-trained model also. ResUnet was developed by Zhengxin Zhang. It has encoder-decoder architecture, with deep residual UNET connections, atrous convolutions, pyramid scene parsing pooling, and multi-tasking inference with a dice loss function. The U-Net uses two 3 x 3 convolution and ReLu activation functions. In the case of ResU-net, these layers are replaced by a pre-activated residual block.

The Convolutional Neural Network has, first two layers have 64 channels of 3*3 filter size and same padding, which have 256X256 size and (3,3) filter size of convolution layers, followed by a max-pooling layer of stride (2, 2) as per same as the previous layer. In deep learning models, the convolution layer should be responsible for the classification and encoder network model for the segmentation. The encoder and decoder models have the skip connection, ReLu activation, and convolution kernels (8-16-32-64-128-256-128-64-32-16-8) for the tested models. Testing and validation of dataset have the loss, batch normalization, and backpropagation for optimization to do testing accuracy with the given models. The current development needs to adopt the deep learning algorithms of self-organ sing map (SOM), multilayer perceptron (MLP), and long short-term memory (LTSM) of deep neural networks from different aspects.
Transfer learning uses previous task knowledge to train the model and is generally used in image classification, segmentation, and prediction of MRI image processing [28]. Figure 1 shows the basic Transfer Learning model of the pre-trained architecture.

## 2. Materials and Methods

Deep learning models are mainly used for object detection and image segmentation. Segmentation and f score are used to improve the accuracy of the pre-trained models. Furthermore, the image segmentation measurement and assessment databases are checked in this article.

### 2.1 Dataset

This dataset from TCIA Archives with MRI images with manual fluid attenuated Flair sequence segmentation masks.

## 3. The Proposed Semantic Models

Semantic segmentation is followed by the classification, the next step is detection and prediction, which provide information on the spatial location of the classes. Finally, semantic segmentation results in the fine-grained inference by making inferring labels and dense predictions for every pixel, which is labeled by the class of its enclosing object and region by the original mask and predicted mask. The proposed Transfer Learning models are added with some more ResNet blocks, Up sample blocks, and Convolution blocks in input layers explained in the Figure 1.

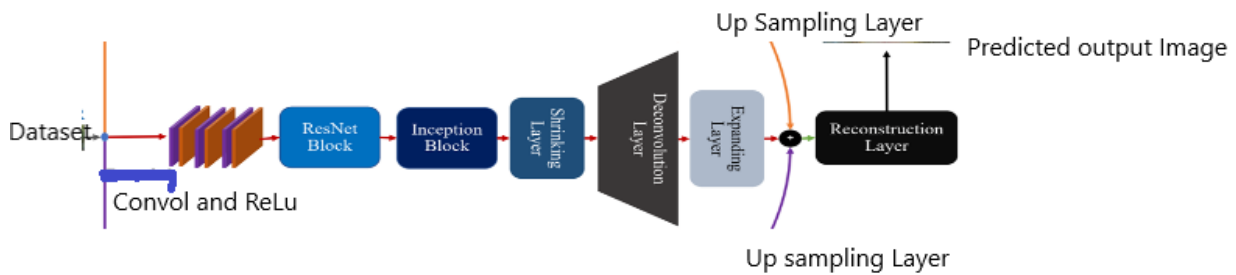

Fig 1: The Proposed Transfer Learning Model

The performance-based approach of DNN architecture [29], VGG-16 model [30], ResNet 50 model [31-32] was discussed in the Table1.

Table 1: Transfer Learning Models used for the Semantic Segmentation and their Limitations

| Pretrained Models | Model description | Advantages | Limitation |
|---|---|---|---|
| **DNN Architecture** | The DNN architecture is a combination of Convolutional Neural Network, Encoder, and Decoder network, | It feeds the input segments of an image to the network, which labels as the pixels | CNN trains the images without sacrificing performance with fewer parameters. DNN needs more parameters, |
| | Regional CNN, and Deep lab V networks. This basic architecture is used to compare with proposed models. | | training is bit slower with LSTM (Long short-term memory). |
| **VGG-16** | VGGNet-16 consists of more filters, and can be trained on GPUs. It has 3X3 kernel depth64, 2X2 max pooling, a fully connected layer, and softmax. | The pre-trained networks for VGG-16 are best for image feature extraction. But this model is limited to the large network architecture weights themselves very slowly | The large network architecture weights themselves very slowly. |
| **ResNet50** | ResNet50 has 48 Convolution layers along with 1 Max pool and 1 Average pool layer. It has 3.8 x 10^9 Floating point operations. The ResNet-50 model was built to support up to 23 million trainable parameters. | ResNet50 suitable for identity-mapping approximation and degradation problems. | ResNet50 requires weeks of training. |
| **ResU-net** | ResU-net model framed with fewer parameters to get high performance. It is an improvement over the existing UNet architecture. | The residual blocks without vanishing gradient or exploding gradients problems and also helps in easy training of the network. | The residual block associated with a squeeze and excitation block. ResU-net used for semantic segmentation problems with class imbalance. |

The ResU-net is much deeper than UNet and VGG16, the model size is fully-connected layers and also reduces the ResNet50 network model size. The pre-trained network can classify more images (such as VGG-16, ResNet50, and ResU-net). Networks with more than thousands of layers can be trained easily without increasing the training error percentage. Fig. 1, shows the different color lines, which represent the multiple landmarks of inputs and the real fundus images. Each blue box corresponds to a convolutional layer, a batch-normal layer, and a ReLu activation and ends up with sigmoid activation. The residual block is associated with a squeeze and excitation block [33]. The ResU-net is used for semantic segmentation problems with class imbalance. Figure 2 shows the proposed segmentation models.

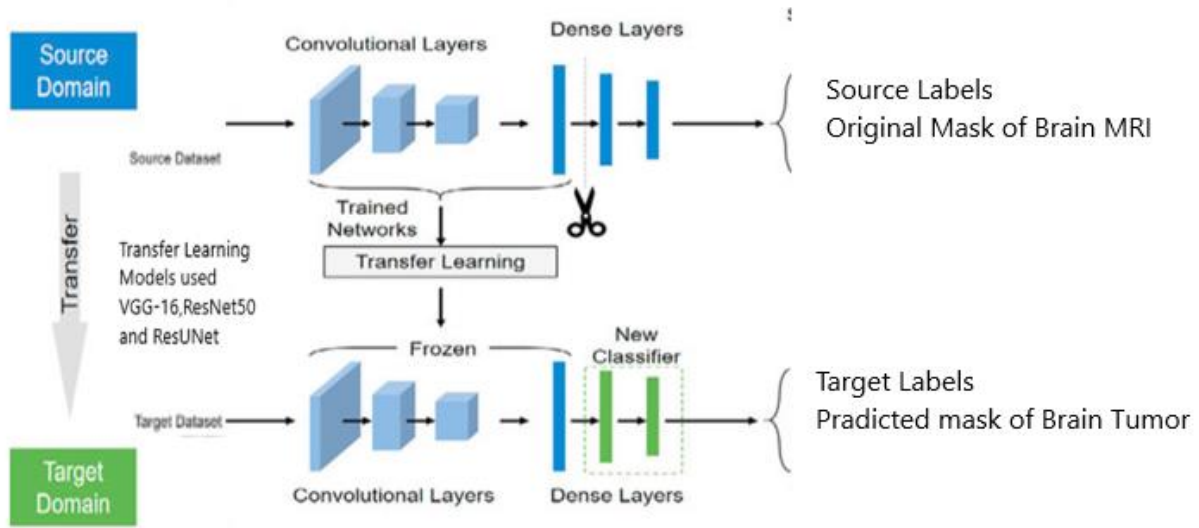

Fig 2: Transfer Learning Models used for the Semantic Segmentation

.
The preprocessed dataset is used for training, validation, and testing process with proposed models of segmentation to find the evaluation metrics. In the image segmentation process, the objects are classified and subdivided to identify the region of interest (RoI) of seizure scars and lesions. A 3D CNN-based fully automated framework has been used [34] to segment the affected MRI images. The trained model uses the deep CNN layers performing 3x3 convolution plus ReLu activation, 2x2 max pooling, creating a high-resolution segmentation map, 2x2 up convolution, overlap strategy of mirroring raw data exploration, augmenting the training data based on the deformed manual labels through no rotation, no shift, no extrapolation and normalization, the loss weight of each pixel produced by training the objects for segmentation mask [35]. The more residual blocks stacking will produce an overfitting problem. To eliminate overfitting the cross-validation, data augmentation, feature selection, regularization, and ensemble methods are used. The Transfer learning models performance is based on the training network, and feed-forward approach to fix the optimized lower level weights and recognize the structures found in images with upper-level retraining weights of back propagation, the model can recognize the features of images, like lesions and tumors, with fewer training examples and less computational power [36].

## 4. Results and Discussion

Initially, the dataset is loaded. images are read from respective directories of Kaggle, along with masks for those images. Then a classification model is built to classify, if a given image has a seizure or not. The basic architecture DNN is used for classification purposes to predict whether the image has a seizure or not. Next, a segmentation model is built to detect the position of the Brain tumor in a given image. Figure 3 explains the performance matrix of the proposed model.

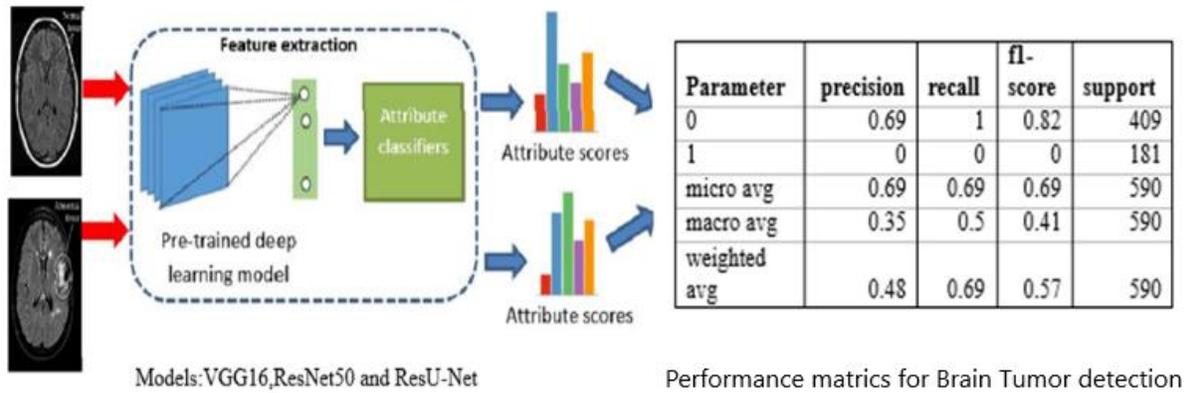

Fig 3: Transfer Learning models for the detection of the Brain Tumor based

The model was built by a Transfer learning models and the dataset of images is trained to predict well and avoid overfitting after 15 epochs through increased batch size. The pretrained convolutional neural network model as a feature extractor. The proposed pre-trained models VGG-16, ResNet50 and ResU-net, are trained on the ImageNet weights to extract features and feed this output to a new classifier to classify and segment the images. The 'ImageNet' weight is used to fetch all models to train on the dataset. The objective is to segment the image, based on the region of interest and features. The models are following these procedures to segment the Brain Tumor. Initially, a data generator is used to test the images and the selected particular model uses the transfer learning and then goes to the function of Freeze the model weights. Second, the model fit is done than the model used to predict the seizure or not, based on the performance indices. Finally, the accuracy score was calculated. Prediction function which takes data frame containing Image ID as input and performs 2 types of image prediction. Initially, whether the image has a defect or not is predicted by classification method, if the model is 99% has no defect, then the image is labeled as no-defect, if the model is not sure, it passes the image to the segmentation network, it again checks if the image has a defect or not to predict the maximum accuracy. Figure 4 shows the basic DNN architecture and proposed model VGG-16 and ResNet50 with mask. Figure 5 explains the Segmented region of Brain Tumor affected predicted mask and original mask. Figure 6 explains about the Brain Tumor of MRI Image, Original mask, Predicted Mask, MRI image with original mask (ground truth) and MRI with predicted Mask of the ResU-net model.

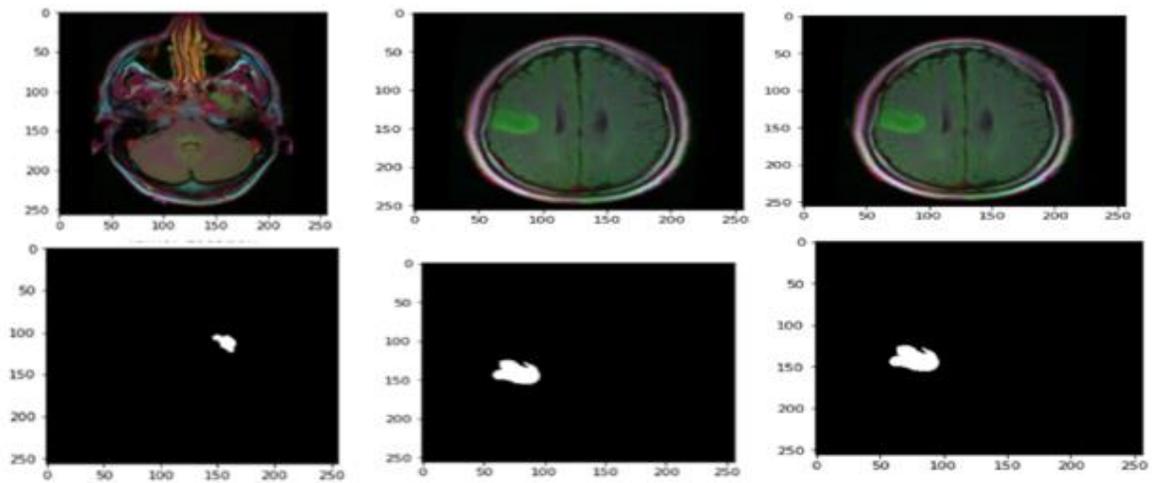

Brain Tumor affected segmented region of MRI images based on the models
(a) DNN Architecture (b) VGG-16 (c) Res Net50

Fig 4: Segmented region of Brain Tumor affected region based on the different models

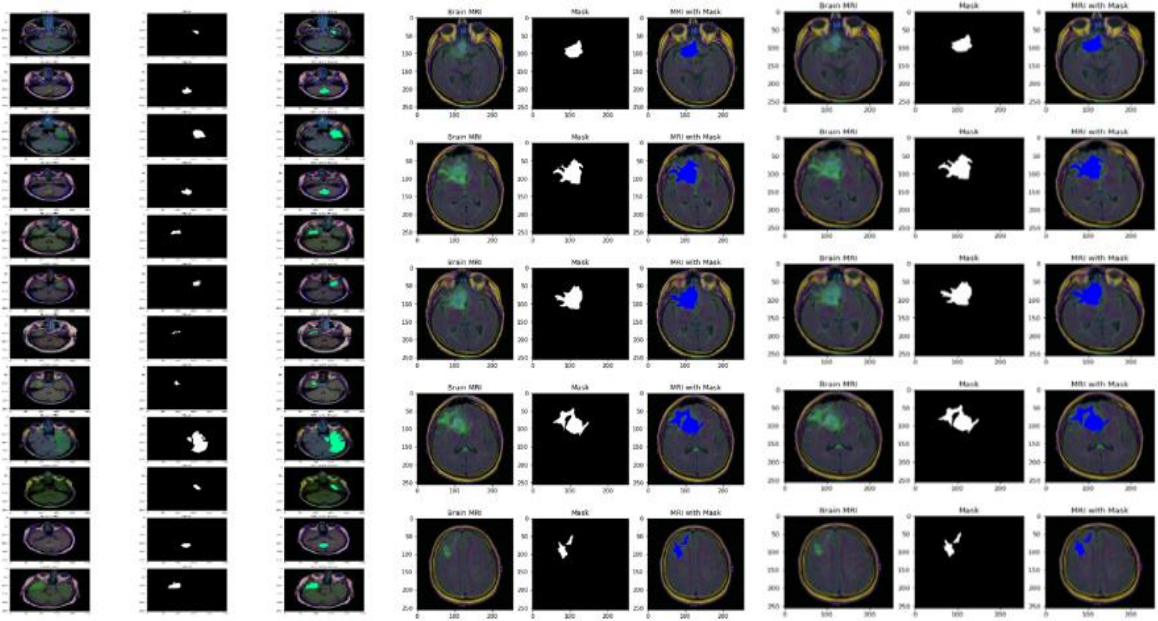

Model :DNN Architecture    Model: VGG-16    Model: ResNet50

Fig 5: Segmented region of Brain Tumor affected predicted mask and original mask based on the different models

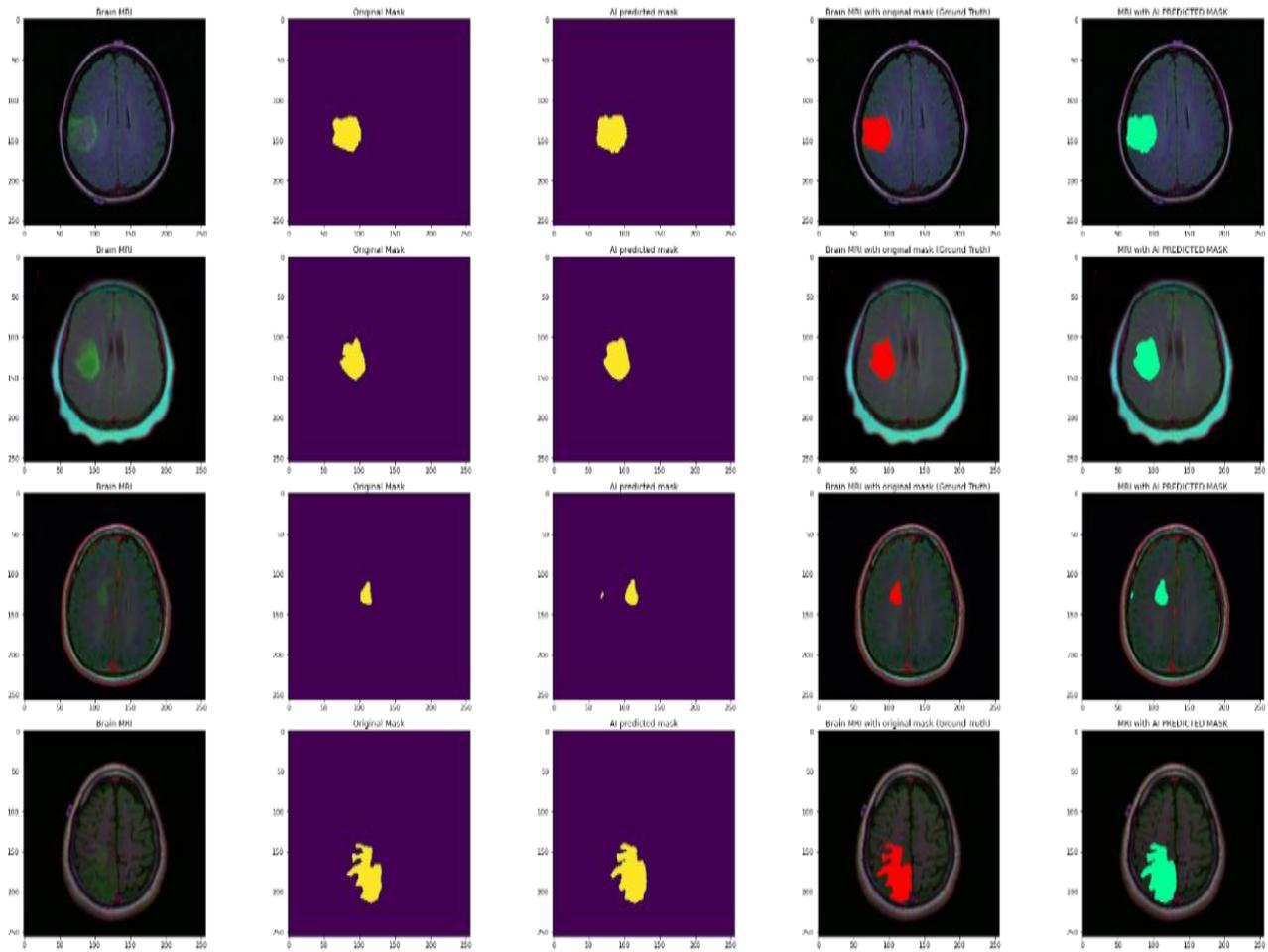

Fig 6: (a). Brain Tumor of MRI Image, (b). Original mask, (c). Predicted Mask, (d). MRI image with original mask (ground truth) a(e) MRI with predicted Mask of the ResU-net model.

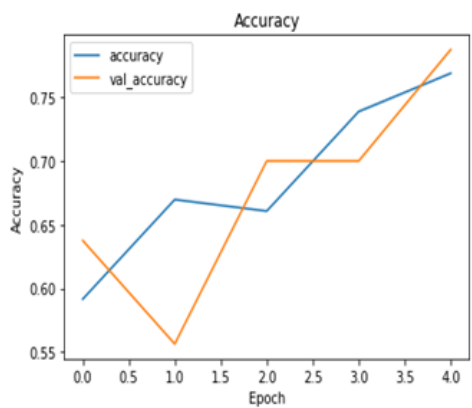
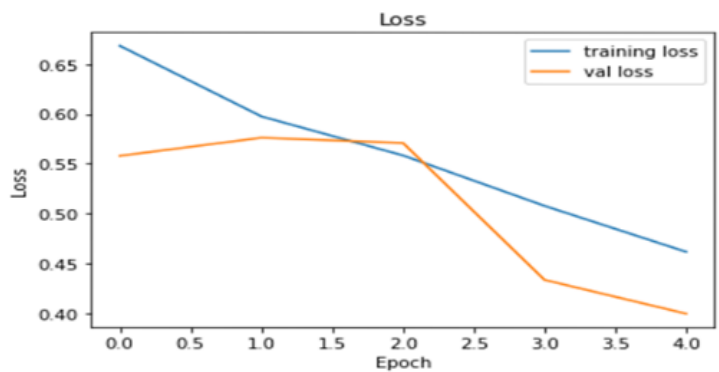

Fig. 7. Training and validation accuracy of model VGG-16 Fig. 8. Training and validation loss of model VGG-16

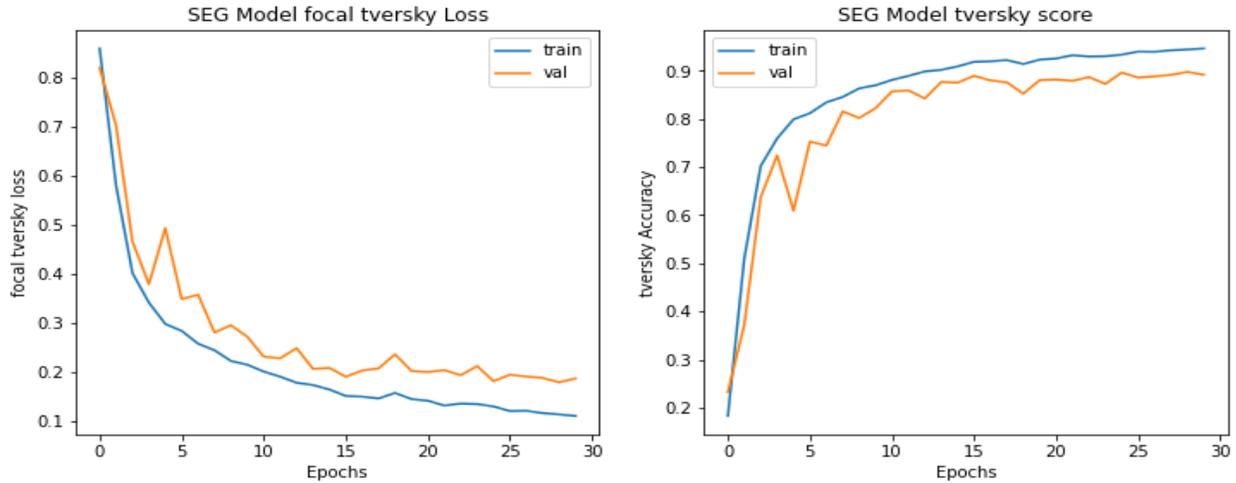

Fig. 9. Training and validation accuracy of model ResUNet Fig. 10. Training and validation loss of model ResU-Net

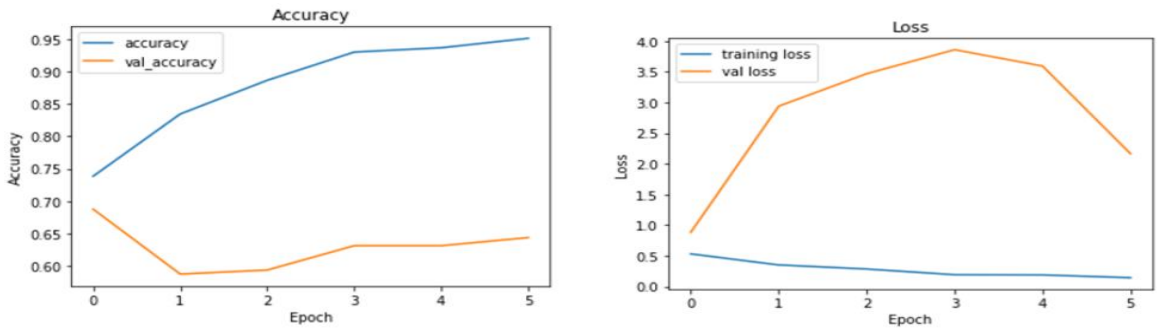

Fig.11. Training and validation accuracy of model ResNet50 Fig. 12. Training and validation loss of model ResNet50

The Figure 7,8,9,10,11 and 12 shows the training and validation accuracy and loss of the semantic models of VGG-16, ResNet50 and ResU-net.

**Table 2:** Models and their ImageNet Trainable parameters

| Proposed Models | Training Accuracy | Validation Accuracy | Training Loss | Validation Loss |
|---|---|---|---|---|
| DNN Architecture | 94.66% | 89.19% | 0.1871 | 0.05 |
| VGG-16 | 76.89% | 78.75% | 0.462 | 0.4 |
| ResU-net | 89.61% | 84.19% | 0.1823 | 0.05 |
| ResNet50 | 95.09% | 68.75% | 0.143 | 0.882 |

Table 2 shows the different semantic segmentation models and their accuracies. The first model DNN builds using a convolutional neural network gives a training accuracy of 94.65 %. The pre-trained model with image augmentation achieved an accuracy of 76.89 % by the VGG-16 model. The pre-trained models (ResU-net) were trained on the dataset of images and fine-tuned with image augmentation to achieve an accuracy of 89.55 %. Lastly, the ResNet 50 model gives a promising result of 95.05%. Table 3 gives the image net trainable values of all the models; Table 4 expresses the models and their performance history.

Table 3: Models and their ImageNet Trainable parameters

| Models | Total Parameters | Trainable Parameters | Non-Trainable Parameters |
|---|---|---|---|
| ResNeT50 | 25,817,218 | 25,764,098 | 53,120 |
| ResUNeT | 1,210,513 | 1,206,129 | 4,384 |
| VGG-16 | 25,817,218 | 25,764,098 | 53,120 |

Table 4: Models and their performance history

| Models | ResNet50 | VGG-16 | ResUNet |
|---|---|---|---|
| Train Images | 2839 | 2839 | 3006 |
| Image-Size | 256x256 | 256x256 | 256x256 |
| Epochs | 15 | 20 | 30 |
| Learning rate | 1.00E-04 | 1.00E-04 | 1.00E-05 |
| ImageNet | 25,764,098 | 25,764,098 | 1,206,129 |
| Time | 155s | 209s | 190s |

For this, VGG-16, ResNet50, and ResU-net architecture are being used to find the training and validation accuracy and loss based on the graph history. The entire architecture is defined and trained as a pre-trained model based on the transfer learning over the dataset. Finally, the results for all three models are combined and then the final image shows the actual mask as well as the predicted mask for each segmented image. The Resnet50 model gives a promising result of accuracy of 95.06%. The transfer learning models, examine the most widely used dataset and their performances, which gives the future research directions in this area.

## 5. Conclusion

The pre-trained model reuse is the concept of Transfer learning. It is used for image classification, segmentation, and prediction. The transfers learning used in this paper are VGG-16, ResU-net and ResNet50. ResNet50 accuracy is more when compared with another trained model with the ImageNet dataset. In future enhancement, transfer learning may be used for real-time clinical datasets of Brain Tumor by glioma-affected image processing to obtain the highest accuracy.